\documentclass[11pt, a4paper]{article}
\pdfoutput=1

\usepackage{amsmath}
\usepackage{amssymb}
\usepackage{comment}
\usepackage{multirow}
\usepackage[utf8]{inputenc}
\usepackage{bm}
\usepackage{cite}
\usepackage{slashed}
\usepackage{braket}
\usepackage{enumitem}
\usepackage{xargs}
\usepackage{tcolorbox}
\usepackage{xfrac}
\usepackage[footnotesize]{caption}
\usepackage{fourier}
\usepackage{cancel}
\usepackage{soul}

\usepackage{ifpdf}
\ifpdf
\usepackage{graphicx, hyperref, xcolor}           
\else                                             
\usepackage[dvipdfmx]{graphicx, hyperref, xcolor} 
\fi
\usepackage{subcaption}

\definecolor{rossoferrari}{HTML}{D9073D}
\definecolor{mediumblue}{HTML}{0000CD}
\definecolor{forestgreen}{HTML}{228B22}
\definecolor{desy_blue}{HTML}{009EE2}
\definecolor{desy_orange}{HTML}{FD8800}
\definecolor{light_pink}{rgb}{1,0.4,0.4}
\definecolor{light_blue}{rgb}{0.284602,0.317763,0.963947}
\hypersetup{
setpagesize=false,
bookmarksnumbered=true,%
bookmarksopen=true,%
colorlinks=true,%
linkcolor=light_blue,
urlcolor=mediumblue,
citecolor=rossoferrari,
linktocpage=false,
}

\newcommand{\revised}[1]{#1}

\usepackage[height=8.85in,width=6.85in]{geometry}
\renewcommand{\baselinestretch}{1.12}

\begin{document}


\begin{titlepage}


~\vspace{-2cm}

\hfill MS-TP-24-13

\hfill DESY-24-098

\begin{center}

\vskip 0.5in

{\LARGE \bf On the overlap reduction function of pulsar timing array\\\smallskip searches for gravitational waves in modified gravity}

\vskip .8in

{\large
Nina Cordes$^{1}$, Andrea Mitridate$^{2}$, Kai Schmitz$^{1,\dagger}$, Tobias Schr\"oder$^{1}$, Kim Wassner$^{1}$
}

\vskip .6in

{\small
\begin{tabular}{ll}
$1$&\!\!\!\!\!\! \emph{University of M\"unster, Institute for Theoretical Physics, 48149 M\"unster, Germany}\\[.3em]
$2$&\!\!\!\!\!\! \emph{Deutsches Elektronen-Synchrotron DESY, Notkestr. 85, 22607 Hamburg, Germany}
\end{tabular}
}
\end{center}


\vskip .6in

\begin{abstract}
Pulsar Timing Array (PTA) searches for gravitational waves (GWs) aim to detect a characteristic correlation pattern in the timing residuals of galactic millisecond pulsars. This pattern is described by the PTA overlap reduction function (ORF) $\Gamma_{ab}\left(\xi_{ab}\right)$, which is known as the Hellings--Downs (HD) curve in general relativity (GR). In theories of modified gravity, the HD curve often receives corrections. Assuming, e.g., a subluminal GW phase velocity, one finds a drastically enhanced ORF in the limit of small angular separations between pulsar $a$ and pulsar $b$ in the sky, $\xi_{ab} \rightarrow 0$. In particular, working in harmonic space and performing an approximate resummation of all multipole contributions, the auto correlation coefficient $\Gamma_{aa}$ seems to diverge. In this paper, we confirm that this divergence is unphysical and provide an exact and analytical expression for $\Gamma_{aa}$ in dependence of the pulsar distance $L_a$ and the GW phase velocity $v_{\rm ph}$. In the GR limit and assuming a large pulsar distance, our expression reduces to $\Gamma_{aa} = 1$. In the case of subluminal phase velocity, we show that the regularization of the naive divergent result is a finite-distance effect, meaning that $\Gamma_{aa}$ scales linearly with $fL_a$, where $f$ is the GW frequency. For superluminal phase velocity (subluminal group velocity), which is relevant in the case of massive gravity, we correct an earlier analytical result for $\Gamma_{ab}$. Our results pave the way for fitting modified-gravity theories with nonstandard phase velocity to PTA data, which requires a proper understanding of the auto correlation coefficient $\Gamma_{aa}$.
\end{abstract}

\vfill

${}^{\dagger}$Corresponding author: \href{mailto:kai.schmitz@uni-muenster.de}{kai.schmitz@uni-muenster.de}


\end{titlepage}


\renewcommand{\thepage}{\arabic{page}}
\setcounter{page}{1}


{\hypersetup{linkcolor=black}\renewcommand{\baselinestretch}{1}\tableofcontents}

\renewcommand{\thepage}{\arabic{page}}
\renewcommand{\thefootnote}{$\natural$\arabic{footnote}}
\setcounter{footnote}{0}


\section{Introduction}

In recent years, the field of pulsar timing array (PTA) searches for gravitational waves (GWs)~\cite{Taylor:2021yjx} has entered an exciting phase, with several PTAs now being on the brink of discovering a stochastic GW background at nHz frequencies~\cite{NANOGrav:2023gor,EPTA:2023fyk,Reardon:2023gzh,Xu:2023wog}. The central observables in PTA measurements are timing residuals for a set of galactic millisecond pulsars, $R_a$, \revised{which correspond to the differences between observed pulse times of arrivals (TOAs) on the one hand and theoretically expected TOAs (according to sophisticated timing models) on the other hand. The timing residuals for each pulsar in the PTA are affected by various noise contributions; however, on top of these noise contributions, GWs can leave an imprint in the $R_a$ in the form of a characteristic correlation pattern. The GW contribution to the timing residuals, $R_a^{\rm GW}$, can notably be written as a time integral over GW-induced redshifts, \revised{$z_a$}, which quantify the instantaneous shift in the pulse TOAs caused by GWs~\cite{Maggiore:2007ulw,Maggiore:2018sht},}
\begin{equation}
R_a^{\rm GW}\left(t\right) = \int_0^t dt' \: z_a\left(t'\right) \,,
\end{equation}
and whose correlator reads
\begin{equation}
\label{eq:zazb}
\left<z_a\left(t\right) z_b\left(t\right)\right> = \frac{2}{3}\int_0^\infty df\:\Gamma_{ab}\left(\xi_{ab},f\right) S_h\left(f\right) \,.
\end{equation}
\revised{Here, $S_h$ is the GW strain power spectrum of the GW Fourier modes $\tilde{h}_A\left(f,\hat{\bm{n}}\right)$ with polarization $A$, frequency $f$, and propagation direction $\hat{\bm{n}}$ that make up the GW background in the nHz frequency band, which we assume to be stochastic, Gaussian, unpolarized, stationary, and isotropic in this paper,}
\begin{equation}
\label{eq:Sh}
\left<\tilde{h}_A\left(f,\hat{\bm{n}}\right)\tilde{h}_{A'}^*\left(f',\hat{\bm{n}}'\right)\right> = \frac{1}{8\pi}\,\delta_{AA'}\,\delta\left(f-f'\right)\,\delta^{(2)}\left(\hat{\bm{n}}-\hat{\bm{n}}'\right)\,S_h\left(f\right) \,, 
\end{equation}
and the function $\Gamma_{ab}$ denotes the PTA overlap reduction function (ORF) (see Ref.~\cite{Romano:2023zhb} for a pedagogical review), 
\begin{equation}
\label{eq:orf}
\Gamma_{ab}\left(\xi_{ab},f\right) = \frac{3}{2}\int \frac{d^2\hat{\bm{n}}}{4\pi}\left[1-e^{-2\pi i f L_a\left(1+\hat{\bm{n}}\cdot\hat{\bm{p}}_a/v_{\rm ph}\right)}\right]\left[1-e^{2\pi i f L_b\left(1+\hat{\bm{n}}\cdot\hat{\bm{p}}_b/v_{\rm ph}\right)}\right] \sum_{A = +,\times} F_a^A\left(\hat{\bm{n}}\right)F_b^A\left(\hat{\bm{n}}\right) \,.
\end{equation}
In this expression, $L_a$ and $L_b$ denote the distances to pulsars $a$ and $b$, respectively; the unit vectors $\hat{\bm{p}}_a$ and $\hat{\bm{p}}_b$ indicate the positions of pulsars $a$ and $b$ on the celestial sphere, respectively; $\xi_{ab}$ is the angular separation of the two pulsars in the sky, $\xi_{ab} = \arccos\left(\hat{\bm{p}}_a\cdot \hat{\bm{p}}_b\right)$; \revised{$v_{\rm ph}$ is the GW phase velocity, which does not necessarily coincide with the speed of light in modified gravity}; and $F_a^+$ and $F_a^\times$ are the PTA detector pattern functions~\cite{Liang:2023ary},
\begin{equation}
\label{eq:FaA}
F_a^A\left(\hat{\bm{n}}\right) = \frac{p_a^i p_a^j e_{ij}^A\left(\hat{\bm{n}}\right)}{2\left(1+\hat{\bm{n}}\cdot\hat{\bm{p}}_a/v_{\rm ph}\right)} \,,
\end{equation}
where $e_{ij}^+$ and $e_{ij}^\times$ are the two GW polarization tensors,
\begin{equation}
\label{eq:epluscross}
e_{ij}^+\left(\hat{\bm{n}}\right) = \hat{\phi}_i\hat{\phi}_j - \hat{\theta}_i\hat{\theta}_j \,, \qquad e_{ij}^\times\left(\hat{\bm{n}}\right) = \hat{\phi}_i\hat{\theta}_j + \hat{\theta}_i\hat{\phi}_j \,,
\end{equation}
which can be constructed in terms of the unit basis vectors $\hat{\bm{\phi}}$ and $\hat{\bm{\theta}}$, 
\begin{equation}
\hat{\bm{\phi}} = \left(\sin \phi,-\cos\phi,0\right)^{\rm T} \,, \qquad \hat{\bm{\theta}} = \left(\cos\theta\cos \phi,\cos\theta\sin\phi,-\sin\theta\right)^{\rm T} \,.
\end{equation}

In this paper, we are interested in the properties of $\Gamma_{ab}$ in theories beyond general relativity (GR)~\cite{Lee:2010cg,Chamberlin:2011ev,Qin:2020hfy,Liang:2021bct,Ezquiaga:2021ler,Hu:2022ujx,Bernardo:2022rif,Schumacher:2023jxq,AnilKumar:2023yfw}. In particular, we shall consider the possibility of a nonstandard GW phase velocity, $v_{\rm ph} \neq 1$, which is realized in various modified-gravity models. At nHz frequencies, GWs may, e.g., exhibit a dispersion relation of the form
\begin{equation}
\omega\left(k\right) = v_{\rm ph}\,k \,, \qquad k = \left|\bm{k}\right| \,, \qquad v_{\rm ph} < 1 \,.
\end{equation}
In this case, the GW phase and group velocities, $v_{\rm ph} = \omega\left(k\right) / k$ and $v_{\rm gr} = \partial \omega\left(k\right) / \partial k$, are identical, and GWs propagate at subluminal speed. Another attractive benchmark consists in massive gravity, where
\begin{equation}
\label{eq:massive}
\omega\left(k\right) = \sqrt{m_g^2 + k^2} \,,
\end{equation}
with $m_g$ denoting the graviton mass. Now, the phase velocity becomes superluminal, while the group velocity remains subluminal. In fact, in the case of massive gravity, the two velocities are just the inverse of each other,
\begin{equation}
v_{\rm ph} = \sqrt{1+\left(m_g/k\right)^2} \,, \qquad v_{\rm gr} = \frac{1}{\sqrt{1+\left(m_g/k\right)^2}} \,.
\end{equation}
In order to be able to fit these two benchmark scenarios of modified gravity to PTA data~\cite{Wu:2023pbt,Bernardo:2023mxc,Wu:2023rib,Bernardo:2023zna,Bi:2023ewq,Liang:2024mex}, it is crucial to have a precise understanding of the ORF, both for separate pulsars at angular separation $\xi_{ab}$ as well as for individual pulsars. Earlier work mostly focused on the cross correlations coefficients $\Gamma_{a\neq b}$. The goal of the present paper therefore is to provide an in-depth discussion of the auto correlation coefficient $\Gamma_{aa}$, including new exact and analytical expressions that are valid both for models with sub- and superluminal phase velocities.

\revised{Metric theories of gravity beyond GR can accommodate up to six GW polarization states~\cite{Eardley:1973zuo,Eardley:1973br}, i.e., two scalar and two vector modes on top of the two tensor modes described by Eq.~\eqref{eq:epluscross}. We expect that much of the machinery that we are going to develop in the present paper can also be applied in a straightforward manner to these non-Einsteinian polarization states, their corresponding ORFs~\cite{Qin:2020hfy}, and in particular their auto correlation coefficients. Nonetheless, we shall follow Ref.~\cite{Liang:2023ary} in this work and restrict ourselves to the two ordinary plus and cross tensor modes, i.e., the type of modes that are generically produced from astrophysical and cosmological sources, whereas other modes are often screened in modified-gravity scenarios~\cite{deRham:2012fw,Chu:2012kz,deRham:2012fg,Joyce:2014kja,Dar:2018dra}.}

The rest of the paper is organized as follows. In the next section, we will briefly review the computation of the ORF in GR, which we will use as an opportunity to fix the overall normalization of the ORF. In Sec.~\ref{sec:Legendre}, we will then turn to the decomposition of the ORF in Legendre polynomials. In the context of this discussion, we will notably encounter a spurious divergence at $\xi_{ab} = 0$ in the case of subluminal $v_{\rm ph}$, which serves as a main motivation for much of the subsequent analysis. Indeed, in Sec.~\ref{sec:auto}, we will present an exact and analytical expression for the auto correlation coefficient $\Gamma_{aa}$, which confirms that the ORF does in fact not diverge. In addition, we will derive a useful and considerably simpler expression for $\Gamma_{aa}$ that manages to approximate the full result with excellent precision at $v_{\rm ph} < 1$. Similarly, in Sec.~\ref{sec:mass}, we will derive a useful and simple expression for $\Gamma_{aa}$ that approximates the full result in the case of massive gravity (i.e., for $v_{\rm ph} > 1$) with excellent precision. Sec.~\ref{sec:conclusions}, finally, contains our conclusions and an outlook on remaining open questions and our next steps.

\section{Hellings--Downs curve}

The two square brackets in Eq.~\eqref{eq:orf} account for the so-called earth-term and pulsar-term contributions to the GW-induced redshifts $z_a$ and $z_b$, respectively. In the GR limit, $v_{\rm ph} \rightarrow 1$, the pulsar-term contributions (i.e., the exponential factors inside the square brackets) yield only fast-oscillating terms that turn out to be numerically negligible when integrated over. In the GR case, one therefore typically only keeps the earth-term contributions (i.e., the factors of $1$ inside the square brackets), such that $\Gamma_{ab}$ obtains the simple form
\begin{equation}
\label{eq:orfgr}
\Gamma_{ab}^{\rm HD}\left(\xi_{ab}\right) \quad\overset{a\neq b}{=}\quad \frac{3}{2}\int \frac{d^2\hat{\bm{n}}}{4\pi} \sum_{A = +,\times} \frac{p_a^ip_a^j e_{ij}^A\left(\hat{\bm{n}}\right)}{2\left(1+\hat{\bm{n}}\cdot\hat{\bm{p}}_a\right)} \frac{p_b^k p_b^l e_{kl}^A\left(\hat{\bm{n}}\right)}{2\left(1+\hat{\bm{n}}\cdot\hat{\bm{p}}_b\right)} \,.
\end{equation}
Note that this expression no longer depends on the GW frequency $f$. The integral in Eq.~\eqref{eq:orfgr} can be evaluated explicitly and results in the Hellings--Downs (HD) curve~\cite{Hellings:1983fr}, the hallmark signature of GWs in PTA data in GR, 
\begin{equation}
\label{eq:HD1}
\Gamma_{ab}^{\rm HD}\left(\xi_{ab}\right) \quad\overset{a\neq b}{=}\quad \frac{3}{2}\,x_{ab} \ln x_{ab} -\frac{x_{ab}}{4} + \frac{1}{2} \,, \qquad x_{ab} = \frac{1}{2}\left(1-\cos \revised{\xi_{ab}}\right) \,.
\end{equation}
The expressions in Eqs.~\eqref{eq:orfgr} and \eqref{eq:HD1} apply to the case of separate pulsars, $a\neq b$, and hence describe the cross correlation between the timing residuals $R_a$ and $R_b$. Starting from Eq.~\eqref{eq:orf}, we can, however, also consider the case of just one pulsar, $a=b$, and compute the auto correlation coefficient $\Gamma_{aa}$. In this case, the correlation between the two pulsar terms is exactly of the same size as the correlation between the two earth terms, 
\begin{equation}
\label{eq:factor2}
\left[1-e^{-2\pi i f L_a\left(1+\hat{\bm{n}}\cdot\hat{\bm{p}}_a\right)}\right]\left[1-e^{2\pi i f L_a\left(1+\hat{\bm{n}}\cdot\hat{\bm{p}}_a\right)}\right] = 2 + \textrm{fast-oscillating terms} \,,
\end{equation}
which means that we need to multiply the HD curve by an extra factor of $2$ in the case of identical pulsars,
\begin{equation}
\label{eq:HD2}
\Gamma_{ab}^{\rm HD}\left(\xi_{ab}\right) = \left(1+\delta_{ab}\right) \left[\frac{3}{2}\,x_{ab} \ln x_{ab} -\frac{x_{ab}}{4} + \frac{1}{2} \right] \,. 
\end{equation}
Furthermore, for identical pulsars, the two position vectors $\hat{\bm{p}}_a$ and $\hat{\bm{p}}_b$ are of course the same, such that $\xi_{ab} = 0$ and $x_{ab} = 0$. In this case, the square brackets in Eq.~\eqref{eq:HD2} yield a factor of $\sfrac{1}{2}$, which, together with the prefactor $1+\delta_{ab}$, results by construction in an auto correlation coefficient of $\Gamma_{aa} = 1$. This convenient normalization of the ORF is a consequence of the factor of $\sfrac{2}{3}$ in Eq.~\eqref{eq:zazb} and the corresponding factor of $\sfrac{3}{2}$ in Eq.~\eqref{eq:orf}. 

\begin{figure}
\begin{center}
\includegraphics[width=0.48\textwidth]{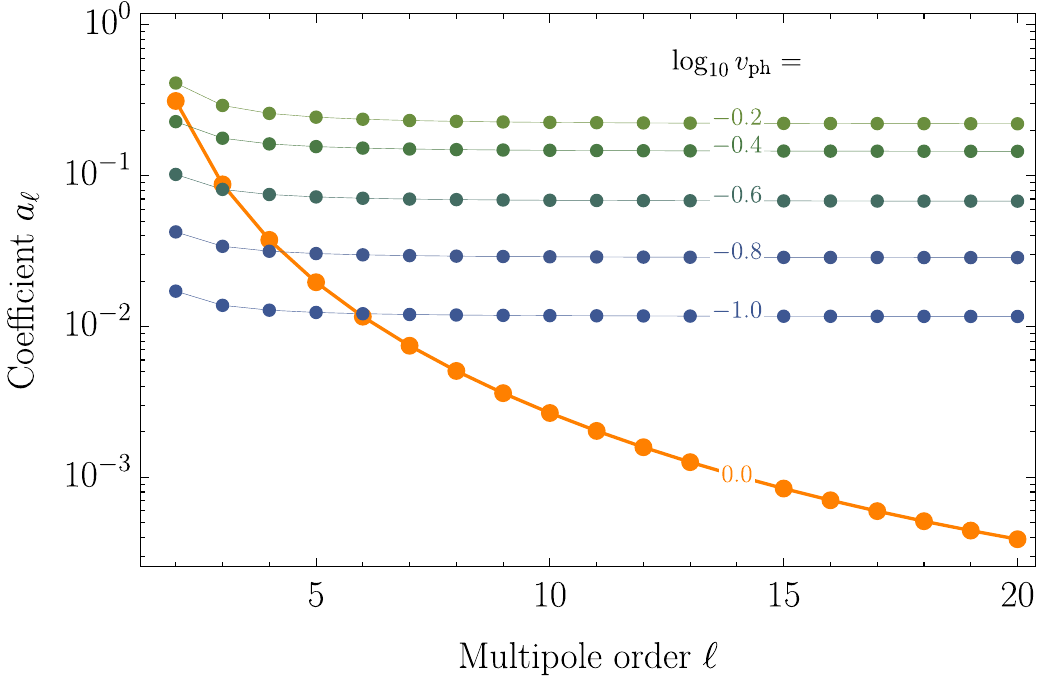}\hfill
\includegraphics[width=0.49\textwidth]{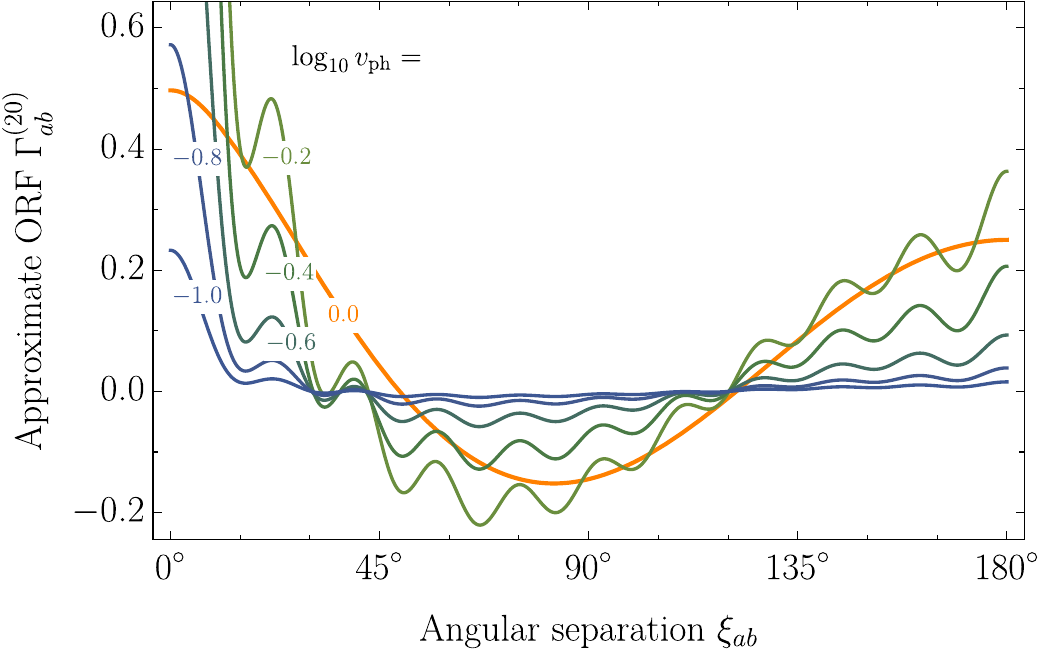}
\end{center}
\caption{Coefficients $a_\ell$ in the Legendre decomposition of the ORF for different values of the GW phase velocity $v_{\rm ph}$, see Eq.~\eqref{eq:Legendre} \textbf{(left panel)}; and the corresponding approximate ORF based on the first $20$ terms in this decomposition, see Eq.~\eqref{eq:ORF20} \textbf{(right panel)}.}
\label{fig:aell}
\end{figure}

\section{Legendre decomposition}
\label{sec:Legendre}

In the more general case of $v_{\rm ph}\neq 1$, it is less straightforward to evaluate the ORF in Eq.~\eqref{eq:orf}. In particular, if $v_{\rm ph} < 1$, it is crucial to retain the pulsar terms in the integrand. Otherwise, the factor $1+\hat{\bm{n}}\cdot\hat{\bm{p}}_a/v_{\rm ph}$ in the denominator of the detector pattern function can result in a spurious numerical singularity. A common approach in the literature therefore consists in decomposing the ORF into Legendre polynomials $P_\ell$~\cite{Gair:2014rwa}, which allows for an approximate numerical evaluation of $\Gamma_{ab}$ order by order in higher multipole contributions, 
\begin{equation}
\label{eq:Legendre}
\Gamma_{ab}\left(\xi_{ab},f\right) = \sum_{\ell = 2}^\infty a_\ell\left(f\right) P_\ell\left(\cos\xi_{ab}\right) \,, \qquad a_\ell\left(f\right) = \frac{3}{2}\left(2\ell+1\right)\,\frac{\left(\ell-2\right)!}{\left(\ell+2\right)!}\frac{\left|c_\ell\left(f\right)\right|^2}{16} \,.
\end{equation}
In GR, the coefficients $c_\ell$ evaluate to $c_\ell = \left(-1\right)^\ell 4$, while in the more general case, we need to compute
\begin{equation}
c_\ell\left(f\right) = \int_{-1}^{+1}dx\left[1 - e^{-i\,2\pi f L\left(1+x/v_{\rm ph}\right)} \right]\frac{\left(1-x^2\right)^2}{1+x/v_{\rm ph}}\,\frac{d^2}{dx^2}P_\ell\left(x\right) \,,
\end{equation}
assuming all pulsars to be located at roughly the same distance, $L_a \sim L_b \sim L$. Keeping only a finite number of terms in the decomposition in Eq.~\eqref{eq:Legendre} will always result in a continuous and smooth approximation of the ORF. By construction, this approach is therefore incapable of encoding the correct auto correlation coefficient $\Gamma_{aa}$, which represents a discontinuous jump in the limit of zero angular separation, $\Gamma_{aa} \neq \lim_{\xi_{ab} \rightarrow 0^+} \Gamma_{ab}$.

For subluminal phase velocity, the coefficients $a_\ell$ decrease less slowly with $\ell$ as in GR (see the left panel of Fig.~\ref{fig:aell}), which results in a slow convergence of the Legendre decomposition in Eq.~\eqref{eq:Legendre}. To see this, consider truncating the expansion in Eq.~\eqref{eq:Legendre} after $\ell = 20$, which results in the following approximate ORF,
\begin{equation}
\label{eq:ORF20}
\Gamma_{ab}^{(20)}\left(\xi_{ab},f\right) = \sum_{\ell = 2}^{20} a_\ell\left(f\right) P_\ell\left(\cos\xi_{ab}\right) \,.
\end{equation}
We plot $\Gamma_{ab}^{(20)}$ for different values of the phase velocity in the right panel of Fig.~\ref{fig:aell}, which illustrates that, for $v_{\rm ph} < 1$, naively summing the first $\mathcal{O}\left(10\right)$ contributions to the Legendre decomposition does not yet yield a satisfactory result. Because of the slow convergence of the coefficients $a_\ell$, the approximate ORF $\Gamma_{ab}^{(20)}$ still features spurious oscillations, which ought to be absent in the full nonperturbative result. 

In order to improve the approximate ORF $\Gamma_{ab}^{(20)}$, it has been proposed to resum the remaining terms in Eq.~\eqref{eq:Legendre}~\cite{Liang:2023ary}, making use of the fact that, for $v_{\rm ph} < 1$, the coefficients $a_\ell$ remain nearly constant at large $\ell$,
\begin{equation}
\Gamma_{ab}^{(\infty)}\left(\xi_{ab},f\right) = \sum_{\ell = 2}^{20} a_\ell\left(f\right) P_\ell\left(\cos\xi_{ab}\right) + \Theta\left(-\log_{10}v\right) a_{20}\left(f\right)\sum_{\ell = 21}^\infty  P_\ell\left(\cos\xi_{ab}\right) \,,
\end{equation}
which can also be written as
\begin{equation}
\label{eq:ORFresummed}
\Gamma_{ab}^{(\infty)}\left(\xi_{ab},f\right) = \sum_{\ell = 2}^{20} a_\ell\left(f\right) P_\ell\left(\cos\xi_{ab}\right) + \Theta\left(-\log_{10}v\right) a_{20}\left(f\right)\left[\frac{1}{\sqrt{2-2\cos\xi_{ab}}} - \sum_{\ell = 0}^{20}  P_\ell\left(\cos\xi_{ab}\right)\right] \,, 
\end{equation}
where we used the following sum rule for the Legendre polynomials, 
\begin{equation}
\sum_{\ell = 0}^\infty  P_\ell\left(\cos\xi_{ab}\right) = \frac{1}{\sqrt{2-2\cos\xi_{ab}}} \,.
\end{equation}
The factor $\Theta$ in Eq.~\eqref{eq:ORFresummed} denotes the Heaviside theta function, i.e., $\Theta\left(x\right) = 1$ for $x > 1$ and $\Theta\left(x\right) = 0$ for $x \leq 0$.

\begin{figure}
\begin{center}
\includegraphics[width=0.75\textwidth]{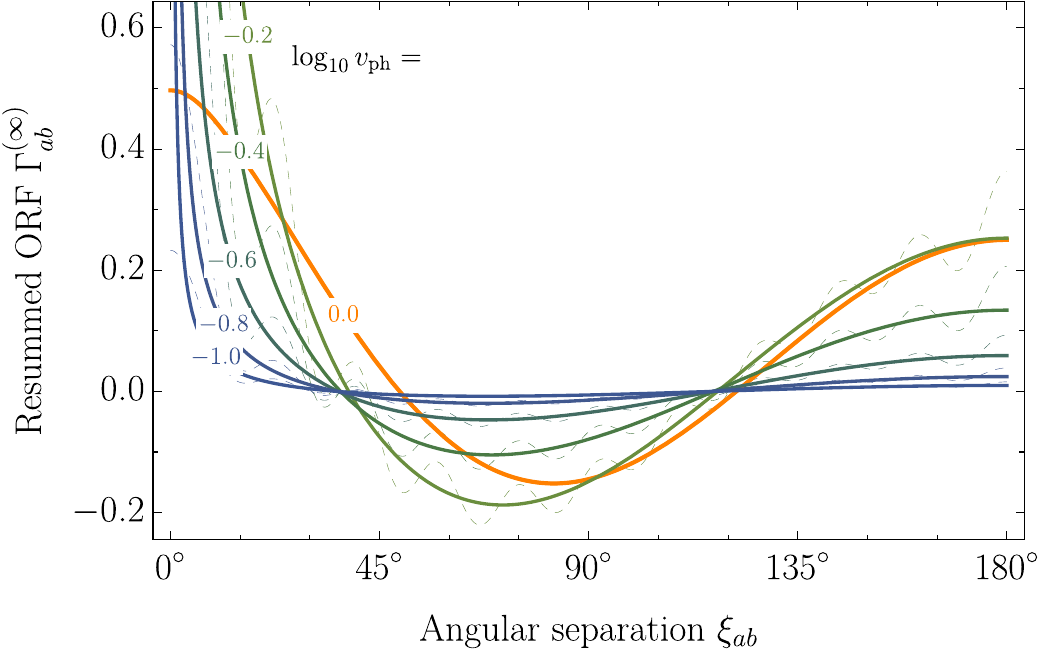}
\end{center}
\caption{Comparison between the resummed ORF $\Gamma_{ab}^{(\infty)}$ in Eq.~\eqref{eq:ORFresummed} (solid lines) and the approximate ORF $\Gamma_{ab}^{(20)}$ in Eq.~\eqref{eq:ORF20} (dashed lines) for different subluminal GW phase velocities $v_{\rm ph}$. The orange line shows $\Gamma_{ab}^{(20)}$ in the GR limit, which requires no improvement.}
\label{fig:ORFresummed}
\end{figure}

The resummed ORF $\Gamma_{ab}^{(\infty)}$ is shown in Fig.~\ref{fig:ORFresummed}, from which it is evident that it significantly improves over the approximate $\Gamma_{ab}^{(20)}$. For practical applications,  $\Gamma_{ab}^{(\infty)}$ thus provides a reasonable approximation of the exact cross correlation coefficients $\Gamma_{ab}$\,---\,at least, at sufficiently large angular separations. In the limit $\xi_{ab} \rightarrow 0$, though, the resummed ORF is unfortunately not well-behaved, since the $\left(2-2\cos\xi_{ab}\right)^{-1/2}$ term in Eq.~\eqref{eq:ORFresummed} causes a divergence at $\xi_{ab} = 0$. This divergence has been known in the literature for some time~\cite{Bi:2023ewq,Liang:2023ary,Liang:2024mex}.%
\footnote{\revised{See also Ref.~\cite{Chamberlin:2011ev} for a discussion of related divergences in the ORFs of other, non-Einsteinian GW polarization states.}}
Still, thus far, it has remained unclear how to regularize the divergence in the resummed ORF, which poses a severe obstacle to using $\Gamma_{ab}^{(\infty)}$ in realistic data analyses. Indeed, fitting modified-gravity models with a nonstandard GW phase velocity to PTA data requires knowledge of the auto correlation coefficients $\Gamma_{aa}$, which are typically (much) larger than the cross correlation coefficients at nonvanishing angular separation, $\Gamma_{a=b} \gtrsim \Gamma_{a\neq b}$. As a consequence, most of the sensitivity to the GW power spectrum $S_h$ in Eq.~\eqref{eq:Sh} derives from measuring the auto correlations of individual pulsars, $\left<R_aR_a\right>$; the information contained in the cross correlations $\left<R_aR_b\right>$, on the other hand, rather serves as a means to confirm the GW nature of the signal. In other words, a successful fit of a modified-gravity model to PTA data must be able to reproduce the expected hierarchy between the auto and cross correlation terms in the correlation matrix $\left<R_aR_b\right>$. Without a proper understanding of the auto correlation coefficients $\Gamma_{aa}$, fit analyses of this type are not feasible.

\begin{figure}
\begin{center}
\includegraphics[width=0.75\textwidth]{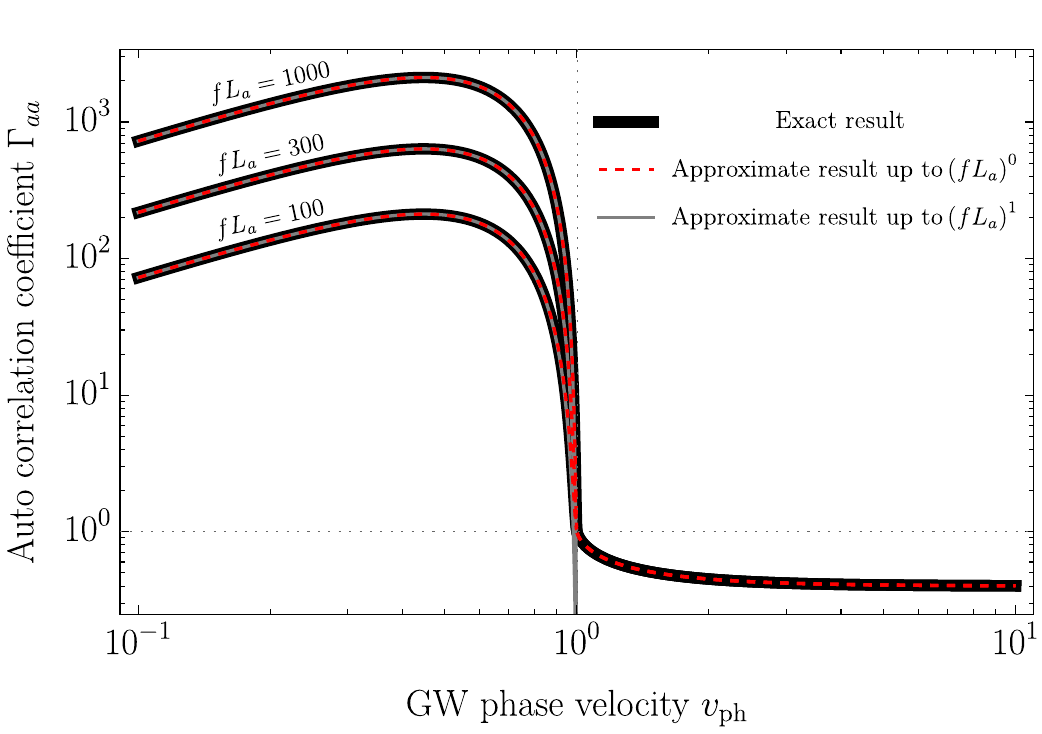}
\end{center}
\caption{Auto correlation coefficient $\Gamma_{aa}$ as a function of the GW phase velocity $v_{\rm ph}$ for three choices of $fL_a$. The thick black lines show the exact analytical result in Eq.~\eqref{eq:Gammaaa}; the red and gray lines refer to the approximate results in Eq.\eqref{eq:ORFNLO} and Eq.~\eqref{eq:ORFLO}, respectively.}
\label{fig:Gammaaa}
\end{figure}

In summary, we conclude that the resummed ORF in Eq.~\eqref{eq:ORFresummed} is not only incapable of resolving the discontinuous jump from $\lim_{\xi_{ab} \rightarrow 0^+} \Gamma_{ab}$ to $\Gamma_{aa}$ at $\xi_{ab} = 0$; it does not even yield a finite result for $\lim_{\xi_{ab} \rightarrow 0^+} \Gamma_{ab}$. The reason for this breakdown of $\Gamma_{ab}^{(\infty)}$ in the zero-$\xi_{ab}$ limit clearly lies in the assumption that the coefficients $a_\ell$ at large $\ell$ can all be approximated by $a_{20}$. If we did not make this assumption and instead summed all exact multipole contributions up to $\ell \rightarrow \infty$, we would expect to obtain a finite result at $\xi_{ab} = 0$. In Ref.~\cite{Liang:2024mex}, the authors speculate in particular that, when pushing the Legendre decomposition to $\ell$ values at least as large as $\ell \sim fL_a$, where $L_a$ is the pulsar distance and $f$ the GW frequency, finite-distance effects should become important and regulate the coefficients $a_\ell$ and hence ultimately lead to a finite ORF at $\xi_{ab} = 0$. In practice, however, such an approach appears infeasible (or at least, less practicable), since $fL_a \sim \mathcal{O}\left(100\cdots 1000\right)$ for typical pulsars.

\section{Auto correlation coefficient}
\label{sec:auto}

In this paper, we will therefore follow a different approach. We shall abandon the decomposition in Legendre  polynomials and the analysis of the ORF in harmonic space and return to the expression for the ORF in Eq.~\eqref{eq:orf}. As it turns out, this expression is perfectly suitable to compute the auto correlation coefficient $\Gamma_{aa}$ in modified-gravity scenarios with $v_{\rm ph}\neq 1$. To do so, let us consider a single pulsar $a$ located in the direction of $\hat{\bm{p}}_a$ in the sky. Without loss of generality, we can always choose our coordinate system such that $\hat{\bm{p}}_a = \left(0,0,1\right)^{\rm T}$. Then, together with $\hat{\bm{n}} = \left(\sin\theta \cos\phi,\sin\theta\sin\phi,\cos\theta\right)^{\rm T}$, the auto correlation coefficient $\Gamma_{aa}$ can be written as
\begin{align}
\Gamma_{aa}\left(f\right) & = \frac{3}{2}\int \frac{d^2\hat{\bm{n}}}{4\pi}\left[1-e^{-2\pi i f L_a\left(1+\hat{\bm{n}}\cdot\hat{\bm{p}}_a/v_{\rm ph}\right)}\right]\left[1-e^{2\pi i f L_a\left(1+\hat{\bm{n}}\cdot\hat{\bm{p}}_a/v_{\rm ph}\right)}\right] \sum_{A = +,\times} \left[\frac{p_a^i p_a^j e_{ij}^A\left(\hat{\bm{n}}\right)}{2\left(1+\hat{\bm{n}}\cdot\hat{\bm{p}}_a/v_{\rm ph}\right)}\right]^2 \\
& = \frac{3}{4}\int_0^\pi d\theta \:\frac{\sin^5\theta\,\sin^2\left(\pi f L_a\left(1+\sfrac{1}{v_{\rm ph}}\cos\theta\right)\right)}{\left(1+\sfrac{1}{v_{\rm ph}}\cos\theta\right)^2} \,. \label{eq:Gammaaa0}
\end{align}
This integral can be solved exactly,
\begin{tcolorbox}
\vspace{-0.5cm}
\begin{align}
& \Gamma_{aa}\left(f\right) = 3\,v_{\rm ph}^4 - 2\,v_{\rm ph}^2 + \frac{v_{\rm ph}^5}{64 y^3} \:\bigg\{\:24\,y\,x_+\,x_- \left[\textrm{Ci}\left(x_+\right) - \textrm{Ci}\left(x_-\right) - 2\,\textrm{arcoth}\left(v_{\rm ph}\right)\right] + 3\,x_+^2 x_-^2 \left[\textrm{Si}\left(x_+\right) - \textrm{Si}\left(x_-\right)\right] \nonumber\\
& + 3\left(2+x_+\,x_-\right)\left(x_- \cos x_+ - x_+ \cos x_-\right) + 6\left[1-\left(3+\sfrac{1}{v_{\rm ph}}\right)y x_-\right] \sin x_+ - 6\left[1-\left(3-\sfrac{1}{v_{\rm ph}}\right)y x_+\right] \sin x_- \:\bigg\}\: \,,
\label{eq:Gammaaa}
\end{align}
\end{tcolorbox}

\noindent
where we introduced the shorthand symbols $x_\pm$ and $y$, and where $\textrm{Ci}$ and $\textrm{Si}$ are the cosine and sine integrals, 
\begin{equation}
x_\pm = \frac{2\,y\left(v_{\rm ph}\pm 1\right)}{v_{\rm ph}} \,, \qquad y = \pi f L_a \,, \qquad \textrm{Ci}\left(x\right) = - \int_x^\infty dt\: \frac{\cos t}{t} \,, \qquad \textrm{Si}\left(x\right) = \int_0^x dt\: \frac{\sin t}{t} \,.
\end{equation}
The auto correlation coefficient $\Gamma_{aa}$ in Eq.~\eqref{eq:Gammaaa} represents the main result of this paper; see Fig.~\ref{fig:Gammaaa} for a plot of $\Gamma_{aa}$ as a function of the GW phase velocity $v_{\rm ph}$ and for three different values of the product $fL_a$.

We shall now discuss some of the properties of the result in Eq.~\eqref{eq:Gammaaa}. First, let us consider the GR limit, 
\begin{equation}
\Gamma_{aa}^{\rm GR}\left(f\right) = \lim_{v_{\rm ph}\rightarrow 1} \Gamma_{aa}\left(f\right) = 1 - \frac{3\left[1 - \textrm{sinc}\left(4\pi fL_a\right)\right]}{8\left(\pi f L_a\right)^2} \,.
\end{equation}
Of course, we could have obtained the same result by simply setting $v_{\rm ph}$ to the speed of light in Eq.~\eqref{eq:Gammaaa0},
\begin{equation}
\label{eq:GammaaaGR}
\Gamma_{aa}^{\rm GR}\left(f\right) = \frac{3}{4}\int_0^\pi d\theta \:\frac{\sin^5\theta\,\sin^2\left(\pi f L_a\left(1+\cos\theta\right)\right)}{\left(1+\cos\theta\right)^2} = 1 - \frac{3\left[1 - \textrm{sinc}\left(4\pi fL_a\right)\right]}{8\left(\pi f L_a\right)^2} \,,
\end{equation}
Interestingly enough, this result for $\Gamma_{aa}^{\rm GR}$ retains the dependence on the GW frequency $f$ and pulsar distance $L_a$. In this sense, $\Gamma_{aa}^{\rm GR}$ differs from the HD result, which just assigns an auto correlation coefficient of $\Gamma_{aa}^{\rm HD} = 1$ to each pulsar. This means that, in principle, the diagonal terms in the correlation matrix $\left<R_a R_b\right>$ do not come with \textit{exactly} the same correlation coefficients. Instead, the different distances to the pulsars in the PTA result in minuscule corrections to the HD result. In practice, however, this effect is safely negligible. Indeed, because PTAs operate in the long-arm limit, we always have $fL_a \gg 1$, such that to excellent approximation,
\begin{equation}
\Gamma_{aa}^{\rm GR}\left(f\right) \quad\overset{fL_a \gg 1}{\approx}\quad \Gamma_{aa}^{\rm HD}\left(f\right) = 1 \,.
\end{equation}
On top, cosmic variance inherent to the HD curve is guaranteed to overshadow any effect of order $\left(fL_a\right)^{-2}$ in the pulsar correlations~\cite{Allen:2022dzg,Allen:2022ksj}. The distinction between $\Gamma_{aa}^{\rm GR}$ and $\Gamma_{aa}^{\rm HD}$ is therefore of little relevance in practice. Nonetheless, we mention that the same result for $\Gamma_{aa}^{\rm GR}$ was already derived in Ref.~\cite{Hu:2022ujx}, which presents a fully analytical expression for the ORF in GR that retains the full dependence on the pulsar terms in Eq.~\eqref{eq:orf}. In the limit of identical pulsars, $\xi_{ab} \rightarrow 0$ and $a=b$, this analytical expressions reduces exactly to the result in Eq.~\eqref{eq:GammaaaGR}.

Next, we observe that the full result in Eq.~\eqref{eq:Gammaaa} can be drastically simplified in the large-$fL_a$ limit. Expanding in inverse powers of $fL_a$, we find that $\Gamma_{aa}$ simply scales linearly with $fL_a$ to leading order, 
\begin{equation}
\label{eq:ORFLO}
\Gamma_{aa}\left(f\right) = \Gamma_{aa}^{\rm LO}\left(f\right) + \mathcal{O}\left(\left(\frac{1}{fL_a}\right)^0\right) \,, \qquad 
\Gamma_{aa}^{\rm LO}\left(f\right) = \Theta\left(-\log_{10}v_{\rm ph}\right) \,\frac{3}{4} \pi^2 fL_a\,  v_{\rm ph} \left(v_{\rm ph}^2-1\right)^2 \,.
\end{equation}
This result for $\Gamma_{aa}^{\rm LO}$ confirms that the divergence that appears in $\Gamma_{aa}^{(\infty)}$ at small angular separations $\xi_{ab}$ and in the regime of subluminal phase velocity $v_{\rm ph}$ is indeed regulated by a finite-distance effect. Moreover, it illustrates that, as soon as $v_{\rm ph}$ is no longer close to the speed of light, such that $(v_{\rm ph}^2-1)^2$ no longer constitutes a limiting suppression factor, $\Gamma_{aa}$ becomes of the order of $fL_a$ and hence parametrically enhanced compared to the cross correlation coefficients $\Gamma_{ab}$ at nonzero angular separations. In order to assess the velocity threshold below which this enhancement becomes relevant, we  solve the condition $\Gamma_{aa}^{\rm LO} \gtrsim 1$ for $v_{\rm ph}$, which results in
\begin{equation}
\label{eq:vphlimit}
v_{\rm ph} \lesssim \bar{v}_{\rm ph} = 1 - \frac{1+\sqrt{3 \pi^2fL_a}}{3 \pi^2 fL_a} \qquad\Rightarrow\qquad \Gamma_{a=b} \gtrsim 1 \gtrsim \Gamma_{a\neq b} \,.
\end{equation}

\begin{figure}
\begin{center}
\includegraphics[width=0.48\textwidth]{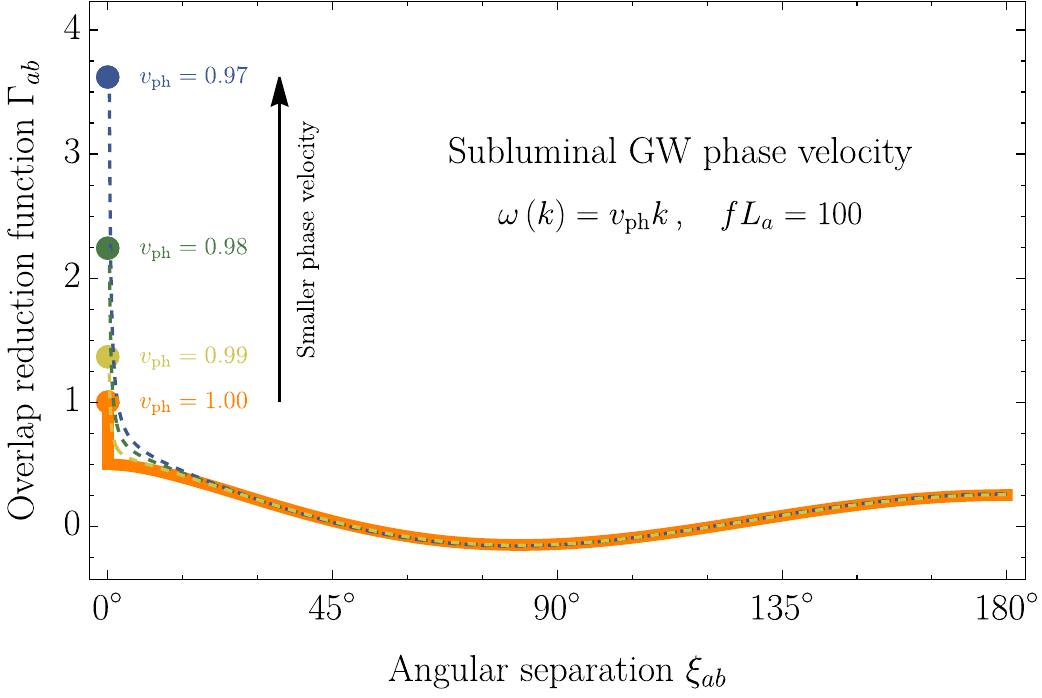}\hfill
\includegraphics[width=0.48\textwidth]{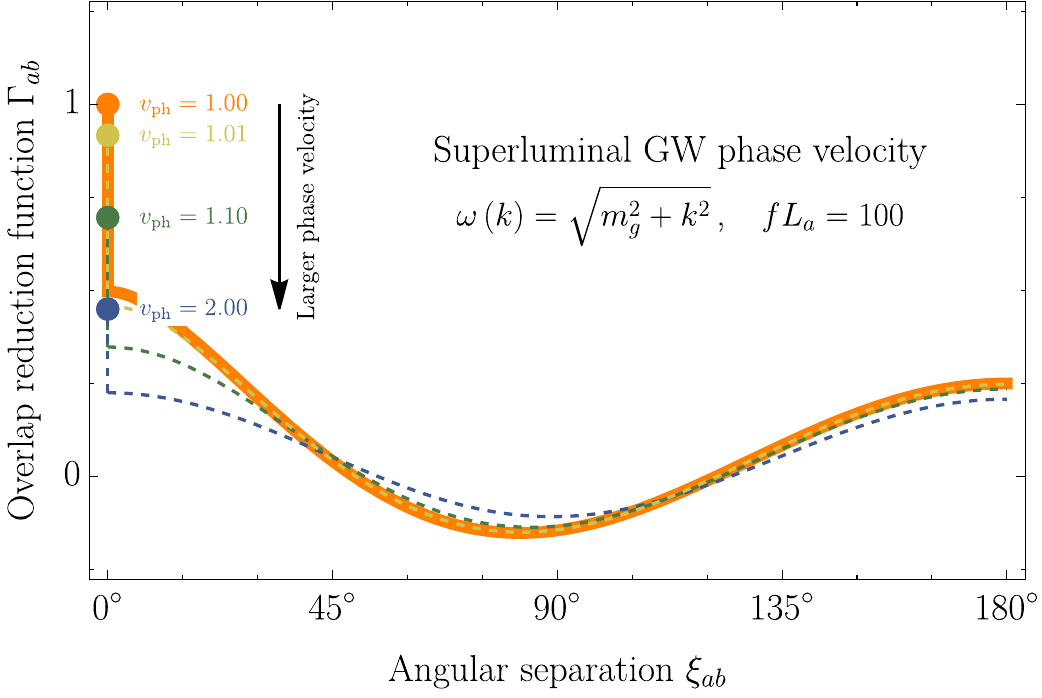}
\end{center}
\caption{Modifications of the HD curve in models with a subluminal \textbf{(left panel)} or superluminal \textbf{(right panel)} GW phase velocity.}
\label{fig:final}
\end{figure}

For typical GW frequencies and pulsar distances, we have $fL_a \sim \mathcal{O}\left(100\cdots 1000\right)$, which means that deviations of $v_{\rm ph}$ from the speed of light by a few percent are enough to cause a large hierarchy between the auto and cross correlations coefficients, as illustrated in the left panel of Fig.~\ref{fig:final}. One must therefore be careful not to be misled by the fact that a decrease in the GW phase velocity by just a few percent leaves much of the ORF at nonzero $\xi_{ab}$ more or less affected. While it is true that, for $1-v_{\rm ph} \sim \mathcal{O}\left(0.01\right)$, the changes in the cross correlations of separate pulsars are practically negligible, it is important to account for the fact that even a small decrease in $v_{\rm ph}$ can drastically increase the ratio $\Gamma_{a=b}/\Gamma_{a\neq b}$. We therefore conclude that the ORF is, in fact, more sensitive to a small decrease in $v_{\rm ph}$ than one may have naively expected; at the smallest possible value of its argument, i.e., at $\xi_{ab} = 0$ and for $a=b$, small changes in $v_{\rm ph}$ can have a big effect. The reason for this enhancement consists in the fact that, for $v_{\rm ph} < 1$, the PTA detector pattern functions in Eq.~\eqref{eq:FaA} become resonantly amplified if the GW propagation direction $\hat{\bm{n}}$ and the pulsar direction $\hat{\bm{p}}_a$ satisfy $\hat{\bm{n}}\cdot\hat{\bm{p}}_a = -v_{\rm ph}$. In GR, this is only possible if $\hat{\bm{n}} = -\hat{\bm{p}}_a$, i.e., for a relative angle between $\hat{\bm{n}}$ and $\hat{\bm{p}}_a$ of $\theta = \pi$. At $\theta = \pi$, the integrand of the integral in Eq.~\eqref{eq:GammaaaGR}, however, vanishes, such that the $\left(1+\cos\theta\right)^2$ term in the denominator has no chance of enhancing the overall result. The situation is different in the case of a subluminal phase velocity, where the $\left(1+\sfrac{1}{v_{\rm ph}}\cos\theta\right)^2$ term in Eq.~\eqref{eq:Gammaaa0} can boost the value of the integrand around $\theta$ values for which $\cos\theta \simeq -v_{\rm ph}$. The propagation direction of the corresponding GWs is thus not antiparallel to the pulsar position, but satisfies instead $\hat{\bm{n}}\cdot\hat{\bm{p}}_a = -v_{\rm ph}$. In passing, we also mention that, despite the fact that the denominator of the PTA detector pattern functions reaches zero at $0 < \theta < \pi$, the integrand in Eq.~\eqref{eq:Gammaaa0} actually never diverges. As discussed at the beginning of Sec.~\ref{sec:Legendre}, the reason for this is that all pulsar terms are kept in deriving Eq.~\eqref{eq:Gammaaa0}. 

The observation that the auto correlation coefficient $\Gamma_{aa}$ becomes strongly enhanced for $v_{\rm ph} < 1$ has important implications for the analysis of PTA data. That is, for $v_{\rm ph}$ (much) smaller than the upper limit in Eq.~\eqref{eq:vphlimit}, it becomes a reasonable approximation to simply neglect all cross correlations and work with
\begin{equation}
v_{\rm ph} \lesssim \bar{v}_{\rm ph} \qquad\Rightarrow\qquad \Gamma_{ab}\left(f\right) \approx \delta_{ab}\,\frac{3}{4} \pi^2 fL_a\,  v_{\rm ph} \left(v_{\rm ph}^2-1\right)^2 \,.
\end{equation}
In the analysis of PTA data, this expression enters the cross power spectrum for the timing residuals $R_a$, which is closely related to the correlator in Eq.~\eqref{eq:zazb} and constructed from the ORF and the strain power spectrum $S_h$,
\begin{equation}
S_{ab}\left(f\right) = \Gamma_{ab}\left(f\right)\, \frac{S_h\left(f\right)}{6\pi^2 f^2} \,.
\end{equation}
Then, if we express $S_h$ in terms of the characteristic strain amplitude, $S_h = h_c^2/\left(2f\right)$, and choose a power-law ansatz for $h_c$, we find the following timing-residual cross power spectrum for modified gravity with $v_{\rm ph} \lesssim \bar{v}_{\rm ph}$,
\begin{equation}
\label{eq:SabApprox}
h_c\left(f\right) = A\left(\frac{f}{f_{\rm ref}}\right)^\alpha \,, \qquad S_{ab}\left(f\right) \approx \delta_{ab}\,fL_a\,\frac{v_{\rm ph} (v_{\rm ph}^2-1)^2A^2}{16 f_{\rm ref}^3}\left(\frac{f}{f_{\rm ref}}\right)^{-\gamma} \,, \qquad \gamma = 3-2\alpha \,.
\end{equation}
This template for $S_{ab}$ shares some similarity with the usual template for a common-spectrum spatially uncorrelated red-noise (CURN) process, in the sense that $S_{ab} \propto \delta_{ab}$. The dependence on $fL_a$, however, represents a departure from a universal ``common spectrum'' for all pulsars. Moreover, the large values of the auto correlation coefficient will cause a suppression in the inferred amplitude $A$ compared to the CURN case. \revised{We plan to confirm this expectation in future work by fitting both our full result for the timing-residual cross power spectrum as well as the approximate result in Eq.~\eqref{eq:SabApprox} to recent PTA data sets.}

\section{Massive gravity}
\label{sec:mass}

Finally, we turn to superluminal phase velocities (subluminal group velocities)\revised{, i.e., to modified-gravity scenarios characterized by the massive dispersion relation in Eq.~\eqref{eq:massive}}. In this case, the leading-order result in Eq.~\eqref{eq:ORFLO}, which simply vanishes at $v_{\rm ph} \geq 1$, no longer suffices. We therefore need to include the next-to-leading order contribution to $\Gamma_{aa}$ when expanding the full result in Eq.~\eqref{eq:Gammaaa} in inverse powers of $fL_a$,
\begin{equation}
\Gamma_{aa}^{\rm NLO}\left(f\right) = \frac{3}{2}\,v_{\rm ph}^3\left(v_{\rm ph}^2-1\right) \ln\left(\frac{\left|v_{\rm ph}-1\right|}{v_{\rm ph}+1}\right) +3\,v_{\rm ph}^4 - 2\,v_{\rm ph}^2 \,.
\end{equation}
Together with the leading-order result in Eq.~\eqref{eq:ORFLO}, we thus obtain
\begin{tcolorbox}
\vspace{-0.5cm}
\begin{align}
\label{eq:ORFNLO}
\Gamma_{aa}\left(f\right) & = \Gamma_{aa}^{\rm LO}\left(f\right) + \Gamma_{aa}^{\rm NLO}\left(f\right) + \mathcal{O}\left(\left(\frac{1}{fL_a}\right)^1\right) \,, \\
\Gamma_{aa}^{\rm LO}\left(f\right) & = \Theta\left(-\log_{10}v_{\rm ph}\right) \,\frac{3}{4} \pi^2 fL_a\,  v_{\rm ph} \left(v_{\rm ph}^2-1\right)^2 \,, \\
\Gamma_{aa}^{\rm NLO}\left(f\right) & = \frac{3}{2}\,v_{\rm ph}^3\left(v_{\rm ph}^2-1\right) \ln\left(\frac{\left|v_{\rm ph}-1\right|}{v_{\rm ph}+1}\right) +3\,v_{\rm ph}^4 - 2\,v_{\rm ph}^2 \,.
\end{align}
\end{tcolorbox}

\noindent In the GR limit, $\Gamma_{aa}^{\rm LO} \rightarrow 0$ and $\Gamma_{aa}^{\rm NLO} \rightarrow 1$, in accord with the fact that $\Gamma_{aa}^{\rm GR} \approx 1$, up to corrections of order $\left(fL_a\right)^{-2}$. The above estimate of the exact auto correlation coefficient in Eq.~\eqref{eq:Gammaaa} is valid at $v_{\rm ph} \geq 1$ and at $v_{\rm ph} < 1$; see Fig.~\ref{fig:Gammaaa}, in which we compare our three results for $\Gamma_{aa}$: the full result in Eq.~\eqref{eq:Gammaaa}, the leading-order expression in Eq.~\eqref{eq:ORFLO}, and the combination of the leading-order and next-to-leading-order terms in Eq.~\eqref{eq:ORFNLO}. In view of this plot, we conclude that $\Gamma_{aa}^{\rm LO}$ serves as an excellent approximation of $\Gamma_{aa}$ at subluminal $v_{\rm ph}$, while the combination $\Gamma_{aa}^{\rm LO} + \Gamma_{aa}^{\rm NLO}$ provides an accurate approximation of $\Gamma_{aa}$ at subluminal and superluminal $v_{\rm ph}$.

In massive gravity, one may prefer to express the dependence of the ORF on the GW speed in terms of the group velocity rather than the phase velocity, $v_{\rm gr} = 1/v_{\rm ph}$. Our result for $\Gamma_{aa}^{\rm NLO}$ in terms of $v_{\rm gr}$ then reads
\begin{equation}
\label{eq:Gammaaavgr}
\Gamma_{aa}^{\rm NLO}\left(f\right) = \frac{1}{2\,v_{\rm gr}^5} \left[6\,v_{\rm gr} - 4\,v_{\rm gr}^3 + 3 \left(v_{\rm gr}^2-1\right) \ln\left(\frac{1+v_{\rm gr}}{1-v_{\rm gr}}\right)\right] \,.
\end{equation}
This expression does not depend on the GW frequency $f$ explicitly, but only implicitly via the group velocity,
\begin{equation}
v_{\rm gr} = \frac{1}{\sqrt{1+\left(m_g/k\right)^2}} = \sqrt{1-\left(\frac{f_g}{f}\right)^2} \,,
\end{equation}
where we used that $\omega= 2\pi f = \sqrt{m_g^2 + k^2}$ in massive gravity, and where $f_g$ is the graviton Compton frequency,
\begin{equation}
f_g = \frac{m_g}{2\pi} \simeq 24.18\,\textrm{nHz}\left(\frac{m_g}{10^{-22}\,\textrm{eV}}\right) \,.
\end{equation}
The result in Eq.~\eqref{eq:Gammaaavgr} can be compared to the known analytical expression for the ORF in massive gravity~\cite{Liang:2021bct,Wu:2023pbt,Wu:2023rib}, which follows from solving the integral in Eq.~\eqref{eq:orf} after discarding the pulsar terms inside the brackets,
\begin{align}
& \Gamma_{a\neq b}^{\rm mass}\left(\xi_{ab},f\right) = \frac{1}{16\,v_{\rm gr}^5}\left[2v_{\rm gr}\left(3+\left(6-5v_{\rm gr}^2\right)\delta\right) - 6\left[1+\delta+v_{\rm gr}^2\left(1-3\delta\right)\right]\ln\left(\frac{1+v_{\rm gr}}{1-v_{\rm gr}}\right)  - \frac{3A}{B}\,\ln C \right] \,, \\
& A = 1+2v_{\rm gr}^2\left(1-2\delta\right)-v_{\rm gr}^4\left(1-2\delta^2\right) \,, \qquad B = \sqrt{\left(1-\delta\right)\left(2-v_{\rm gr}^2\left(1+\delta\right)\right)} \,, \qquad C = \frac{A-2v_{\rm gr}\left(1-v_{\rm gr}^2\delta\right)B}{\left(v_{\rm gr}^2-1\right)^2} \,,
\end{align}
where the dependence on the angular separation is encoded in $\delta =\cos\xi_{ab}$. In the limit $\xi_{ab}\rightarrow 0$, we find
\begin{equation}
\lim_{\xi_{ab}\rightarrow 0^+} \Gamma_{a\neq b}^{\rm mass}\left(\xi_{ab},f\right) = \frac{1}{4\,v_{\rm gr}^5} \left[6\,v_{\rm gr} - 4\,v_{\rm gr}^3 + 3 \left(v_{\rm gr}^2-1\right) \ln\left(\frac{1+v_{\rm gr}}{1-v_{\rm gr}}\right)\right] = \frac{1}{2}\,\Gamma_{aa}^{\rm NLO}\left(f\right) \,,
\end{equation}
which equals exactly one half of our result for $\Gamma_{aa}^{\rm NLO}$. We thus conclude that the full expression for the ORF in massive gravity, valid for both separate and identical pulsars, requires the same $1+\delta_{ab}$ prefactor as in GR,
\begin{tcolorbox}
\vspace{-0.3cm}
\begin{equation}
\Gamma_{ab}^{\rm mass}\left(\xi_{ab},f\right) = \frac{1+\delta_{ab}}{16\,v_{\rm gr}^5}\left[2v_{\rm gr}\left(3+\left(6-5v_{\rm gr}^2\right)\delta\right) - 6\left[1+\delta+v_{\rm gr}^2\left(1-3\delta\right)\right]\ln\left(\frac{1+v_{\rm gr}}{1-v_{\rm gr}}\right)  - \frac{3A}{B}\,\ln C \right] \,.
\end{equation}
\end{tcolorbox}

\noindent
To the best of our knowledge, the prefactor of $1+\delta_{ab}$ has thus far been overlooked in earlier studies. Just like in GR, the extra factor of $2$ for $a=b$ originates from the auto correlation of the pulsar term in Eq.~\eqref{eq:factor2}. 

\section{Conclusions}
\label{sec:conclusions}

PTA searches for GWs require a precise understanding of the ORF $\Gamma_{ab}$, which is sensitive to the presence of new physics in the gravity sector. For instance, the HD curve in GR receives corrections if the GW phase velocity $v_{\rm ph}$ does not coincide with the speed of light. In the literature, the ORF had been previously calculated for such scenarios of modified gravity for both for $v_{\rm ph} < 1$ and $v_{\rm ph} > 1$ by various authors. Most existing analyses, however, only consider the case of separate pulsars ($a\neq b$) and neglect the case of identical pulsars ($a=b$). In this paper, we therefore revisited the computation of the ORF in scenarios with a nonstandard phase velocity, supplementing the known expressions for the cross correlation coefficients $\Gamma_{a\neq b}$ by a careful analytical evaluation of the auto correlation coefficients $\Gamma_{a=b}$; see Fig.~\eqref{fig:final} for a visual illustration of our main results.

In the GR limit, $v_{\rm ph} \rightarrow 1$, we recover the standard HD result, $\Gamma_{aa}^{\rm HD} = 1$, at leading order, but also find (numerically negligible) subleading corrections of the order of $(fL_a)^{-2}$, which introduce (at least, in principle) a dependence on the pulsar distance $L_a$ in the diagonal entries of $\left<R_a R_b\right>$. For subluminal $v_{\rm ph}$, we confirm that the divergence of the ORF in the limit of small angular separations that had been observed in earlier works is unphysical and instead regulated by the finite distances to the pulsars in the PTA.  At leading order, we notably find that the auto correlation coefficient scales linearly with $fL_a$; see Eq.~\eqref{eq:ORFLO}. We therefore conclude that, for a sufficiently small phase velocity, $\Gamma_{a=b}$ becomes parametrically enhanced compared to $\Gamma_{a\neq b}$, which means that the cross correlations among pulsars become insignificant to good approximation in this model.

Meanwhile, we remark that the known expression for $\Gamma_{ab}$ in Eq.~\eqref{eq:ORFresummed} still suffers from a certain degree of uncertainty at very small values of $\xi_{ab}$. The point is: While we derived the finite auto correlation coefficient for subluminal $v_{\rm ph}$ from first principles, we did not study how the approximate resummation of higher multipole terms in the Legendre decomposition (or any other way of computing the ORF, for that matter) needs to be modified such that $\Gamma_{ab}$ does indeed approach $\Gamma_{aa}$ in the limit $\xi_{ab}\rightarrow 0$ and for $a = b$. In other words, our calculation tells us the value that the ORF must take at $\xi_{ab} = 0$ and for $a=b$; we, however, still do not have a complete understanding of how this value is actually reached as $\xi_{ab} \rightarrow 0$ and $L_a \rightarrow L_b$. For instance, in GR and for superluminal $v_{\rm ph}$, we know that the ORF first reaches a finite value, $\lim_{\xi_{ab}\rightarrow 0^+}\Gamma_{ab}$, which then needs to be multiplied by a factor of $1+\delta_{ab}$, in order to produce the final value of $\Gamma_{aa}$. \revised{Moreover, a careful analysis of the behavior of the full ORF in GR in the double limit $\xi_{ab}\rightarrow 0$ and $L_a \rightarrow L_b$ was performed in Ref.~\cite{Mingarelli:2014xfa}.} For subluminal $v_{\rm ph}$, on the other hand, the exact relation between $\lim_{\xi_{ab}\rightarrow 0^+}\Gamma_{ab}$ and $\Gamma_{aa}$ remains unclear at present. For practical applications, this limitation is of little relevance, though, as long as the angular separations between all pulsars in the PTA are bounded from below. We checked, e.g., that the expression in Eq.~\eqref{eq:ORFresummed} is insensitive to changes in $fL_a$ for angular separations of at least $\xi_{ab} = \pi/100$. We therefore expect that finite-distance effects can be neglected for all but the tiniest angular separations, which, however, are not realized in typical PTAs. In summary, this means that, in models with $v_{\rm ph} < 1$ and for PTAs in which no two pulsars are extremely close together in the sky, it is justified to use our result for $\Gamma_{aa}$ to describe the auto correlations of individual pulsars, in combination with Eq.~\eqref{eq:ORFresummed} to compute the cross correlations among pulsars. At the same time, the exact relation between $\lim_{\xi_{ab}\rightarrow 0^+}\Gamma_{ab}$ and $\Gamma_{aa}$ for $v_{\rm ph} < 1$ remains a relevant conceptual question for future work; \revised{see Ref.~\cite{Domenech:2024pow} for recent progress in this direction}. \revised{Similarly, it would be interesting to extend the results that we obtained in the present paper to non-Einsteinian polarization modes.}

Finally, for superluminal $v_{\rm ph}$, which is relevant in the case of massive gravity, we obtain the known analytical result for $\Gamma_{aa}$, up to a factor of $2$, which had been overlooked in earlier studies. In conclusion, the results presented in this paper now pave the way for fitting modified-gravity models with sub- or superluminal phase velocity to PTA data. We will present the result of such fits to recent PTA data in an upcoming publication.


\subsection*{Acknowledgments}


We thank Qiuyue Liang and Mark Trodden for helpful discussions.
This work was supported by the Deutsche Forschungsgemeinschaft under Germany's Excellence Strategy -- EXC 2121 Quantum Universe -- 390833306. 
The work of K.\,S.\ and T.\,S.\ is supported by Deutsche Forschungsgemeinschaft (DFG) through the Research Training Group (Graduiertenkolleg) 2149: Strong and Weak Interactions -- from Hadrons to Dark Matter.


\small
\bibliographystyle{utphys}
\bibliography{arXiv_2}

\providecommand{\href}[2]{#2}\begingroup\raggedright\begin{thebibliography}{10}

\bibitem{Taylor:2021yjx}
S.~R. Taylor, ``{The Nanohertz Gravitational Wave Astronomer},''
  \href{http://arxiv.org/abs/2105.13270}{{\ttfamily arXiv:2105.13270
  [astro-ph.HE]}}.

\bibitem{NANOGrav:2023gor}
{\bfseries NANOGrav} Collaboration, G.~Agazie {\em et~al.}, ``{The NANOGrav 15
  yr Data Set: Evidence for a Gravitational-wave Background},''
  \href{http://dx.doi.org/10.3847/2041-8213/acdac6}{{\em Astrophys. J. Lett.}
  {\bfseries 951} no.~1, (2023) L8},
  \href{http://arxiv.org/abs/2306.16213}{{\ttfamily arXiv:2306.16213
  [astro-ph.HE]}}.

\bibitem{EPTA:2023fyk}
{\bfseries EPTA, InPTA:} Collaboration, J.~Antoniadis {\em et~al.}, ``{The
  second data release from the European Pulsar Timing Array - III. Search for
  gravitational wave signals},''
  \href{http://dx.doi.org/10.1051/0004-6361/202346844}{{\em Astron. Astrophys.}
  {\bfseries 678} (2023) A50},
  \href{http://arxiv.org/abs/2306.16214}{{\ttfamily arXiv:2306.16214
  [astro-ph.HE]}}.

\bibitem{Reardon:2023gzh}
D.~J. Reardon {\em et~al.}, ``{Search for an Isotropic Gravitational-wave
  Background with the Parkes Pulsar Timing Array},''
  \href{http://dx.doi.org/10.3847/2041-8213/acdd02}{{\em Astrophys. J. Lett.}
  {\bfseries 951} no.~1, (2023) L6},
  \href{http://arxiv.org/abs/2306.16215}{{\ttfamily arXiv:2306.16215
  [astro-ph.HE]}}.

\bibitem{Xu:2023wog}
H.~Xu {\em et~al.}, ``{Searching for the Nano-Hertz Stochastic Gravitational
  Wave Background with the Chinese Pulsar Timing Array Data Release I},''
  \href{http://dx.doi.org/10.1088/1674-4527/acdfa5}{{\em Res. Astron.
  Astrophys.} {\bfseries 23} (2023) 075024},
  \href{http://arxiv.org/abs/2306.16216}{{\ttfamily arXiv:2306.16216
  [astro-ph.HE]}}.

\bibitem{Maggiore:2007ulw}
M.~Maggiore, {\em {Gravitational Waves. Vol. 1: Theory and Experiments}}.
\newblock Oxford University Press, 2007.

\bibitem{Maggiore:2018sht}
M.~Maggiore, {\em {Gravitational Waves. Vol. 2: Astrophysics and Cosmology}}.
\newblock Oxford University Press, 3, 2018.

\bibitem{Romano:2023zhb}
J.~D. Romano and B.~Allen, ``{Answers to frequently asked questions about the
  pulsar timing array Hellings and Downs curve},''
  \href{http://arxiv.org/abs/2308.05847}{{\ttfamily arXiv:2308.05847 [gr-qc]}}.

\bibitem{Liang:2023ary}
Q.~Liang, M.-X. Lin, and M.~Trodden, ``{A test of gravity with Pulsar Timing
  Arrays},'' \href{http://dx.doi.org/10.1088/1475-7516/2023/11/042}{{\em JCAP}
  {\bfseries 11} (2023) 042}, \href{http://arxiv.org/abs/2304.02640}{{\ttfamily
  arXiv:2304.02640 [astro-ph.CO]}}.

\bibitem{Lee:2010cg}
K.~Lee, F.~A. Jenet, R.~H. Price, N.~Wex, and M.~Kramer, ``{Detecting massive
  gravitons using pulsar timing arrays},''
  \href{http://dx.doi.org/10.1088/0004-637X/722/2/1589}{{\em Astrophys. J.}
  {\bfseries 722} (2010) 1589--1597},
  \href{http://arxiv.org/abs/1008.2561}{{\ttfamily arXiv:1008.2561
  [astro-ph.HE]}}.

\bibitem{Chamberlin:2011ev}
S.~J. Chamberlin and X.~Siemens, ``{Stochastic backgrounds in alternative
  theories of gravity: overlap reduction functions for pulsar timing arrays},''
  \href{http://dx.doi.org/10.1103/PhysRevD.85.082001}{{\em Phys. Rev. D}
  {\bfseries 85} (2012) 082001},
  \href{http://arxiv.org/abs/1111.5661}{{\ttfamily arXiv:1111.5661
  [astro-ph.HE]}}.

\bibitem{Qin:2020hfy}
W.~Qin, K.~K. Boddy, and M.~Kamionkowski, ``{Subluminal stochastic
  gravitational waves in pulsar-timing arrays and astrometry},''
  \href{http://dx.doi.org/10.1103/PhysRevD.103.024045}{{\em Phys. Rev. D}
  {\bfseries 103} (2021) 024045},
  \href{http://arxiv.org/abs/2007.11009}{{\ttfamily arXiv:2007.11009 [gr-qc]}}.

\bibitem{Liang:2021bct}
Q.~Liang and M.~Trodden, ``{Detecting the stochastic gravitational wave
  background from massive gravity with pulsar timing arrays},''
  \href{http://dx.doi.org/10.1103/PhysRevD.104.084052}{{\em Phys. Rev. D}
  {\bfseries 104} (2021) 084052},
  \href{http://arxiv.org/abs/2108.05344}{{\ttfamily arXiv:2108.05344
  [astro-ph.CO]}}.

\bibitem{Ezquiaga:2021ler}
J.~M. Ezquiaga, W.~Hu, M.~Lagos, and M.-X. Lin, ``{Gravitational wave
  propagation beyond general relativity: waveform distortions and echoes},''
  \href{http://dx.doi.org/10.1088/1475-7516/2021/11/048}{{\em JCAP} {\bfseries
  11} no.~11, (2021) 048}, \href{http://arxiv.org/abs/2108.10872}{{\ttfamily
  arXiv:2108.10872 [astro-ph.CO]}}.

\bibitem{Hu:2022ujx}
Y.~Hu, P.-P. Wang, Y.-J. Tan, and C.-G. Shao, ``{Full analytic expression of
  overlap reduction function for gravitational wave background with pulsar
  timing arrays},'' \href{http://dx.doi.org/10.1103/PhysRevD.106.024005}{{\em
  Phys. Rev. D} {\bfseries 106} no.~2, (2022) 024005},
  \href{http://arxiv.org/abs/2205.09272}{{\ttfamily arXiv:2205.09272 [gr-qc]}}.

\bibitem{Bernardo:2022rif}
R.~C. Bernardo and K.-W. Ng, ``{Stochastic gravitational wave background
  phenomenology in a pulsar timing array},''
  \href{http://dx.doi.org/10.1103/PhysRevD.107.044007}{{\em Phys. Rev. D}
  {\bfseries 107} (2023) 044007},
  \href{http://arxiv.org/abs/2208.12538}{{\ttfamily arXiv:2208.12538 [gr-qc]}}.

\bibitem{Schumacher:2023jxq}
K.~Schumacher, N.~Yunes, and K.~Yagi, ``{Gravitational wave polarizations with
  different propagation speeds},''
  \href{http://dx.doi.org/10.1103/PhysRevD.108.104038}{{\em Phys. Rev. D}
  {\bfseries 108} no.~10, (2023) 104038},
  \href{http://arxiv.org/abs/2308.05589}{{\ttfamily arXiv:2308.05589 [gr-qc]}}.

\bibitem{AnilKumar:2023yfw}
N.~Anil~Kumar and M.~Kamionkowski, ``{All the Pretty Overlap Reduction
  Functions},'' \href{http://arxiv.org/abs/2311.14159}{{\ttfamily
  arXiv:2311.14159 [astro-ph.CO]}}.

\bibitem{Wu:2023pbt}
Y.-M. Wu, Z.-C. Chen, and Q.-G. Huang, ``{Search for stochastic
  gravitational-wave background from massive gravity in the NANOGrav 12.5-year
  dataset},'' \href{http://dx.doi.org/10.1103/PhysRevD.107.042003}{{\em Phys.
  Rev. D} {\bfseries 107} no.~4, (2023) 042003},
  \href{http://arxiv.org/abs/2302.00229}{{\ttfamily arXiv:2302.00229 [gr-qc]}}.

\bibitem{Bernardo:2023mxc}
R.~C. Bernardo and K.-W. Ng, ``{Constraining gravitational wave propagation
  using pulsar timing array correlations},''
  \href{http://dx.doi.org/10.1103/PhysRevD.107.L101502}{{\em Phys. Rev. D}
  {\bfseries 107} no.~10, (2023) L101502},
  \href{http://arxiv.org/abs/2302.11796}{{\ttfamily arXiv:2302.11796 [gr-qc]}}.

\bibitem{Wu:2023rib}
Y.-M. Wu, Z.-C. Chen, Y.-C. Bi, and Q.-G. Huang, ``{Constraining the graviton
  mass with the NANOGrav 15\,year data set},''
  \href{http://dx.doi.org/10.1088/1361-6382/ad2a9b}{{\em Class. Quant. Grav.}
  {\bfseries 41} no.~7, (2024) 075002},
  \href{http://arxiv.org/abs/2310.07469}{{\ttfamily arXiv:2310.07469
  [astro-ph.CO]}}.

\bibitem{Bernardo:2023zna}
R.~C. Bernardo and K.-W. Ng, ``{Beyond the Hellings-Downs curve:
  Non-Einsteinian gravitational waves in pulsar timing array correlations},''
  \href{http://arxiv.org/abs/2310.07537}{{\ttfamily arXiv:2310.07537 [gr-qc]}}.

\bibitem{Bi:2023ewq}
Y.-C. Bi, Y.-M. Wu, Z.-C. Chen, and Q.-G. Huang, ``{Constraints on the velocity
  of gravitational waves from the NANOGrav 15-year data set},''
  \href{http://dx.doi.org/10.1103/PhysRevD.109.L061101}{{\em Phys. Rev. D}
  {\bfseries 109} no.~6, (2024) L061101},
  \href{http://arxiv.org/abs/2310.08366}{{\ttfamily arXiv:2310.08366
  [astro-ph.CO]}}.

\bibitem{Liang:2024mex}
Q.~Liang, I.~Obata, and M.~Sasaki, ``{Testing Gravity with Frequency-Dependent
  Overlap Reduction Function in Pulsar Timing Array},''
  \href{http://arxiv.org/abs/2405.11755}{{\ttfamily arXiv:2405.11755
  [astro-ph.CO]}}.

\bibitem{Eardley:1973zuo}
D.~M. Eardley, D.~L. Lee, and A.~P. Lightman, ``{Gravitational-wave
  observations as a tool for testing relativistic gravity},''
  \href{http://dx.doi.org/10.1103/PhysRevD.8.3308}{{\em Phys. Rev. D}
  {\bfseries 8} (1973) 3308--3321}.

\bibitem{Eardley:1973br}
D.~M. Eardley, D.~L. Lee, A.~P. Lightman, R.~V. Wagoner, and C.~M. Will,
  ``{Gravitational-wave observations as a tool for testing relativistic
  gravity},'' \href{http://dx.doi.org/10.1103/PhysRevLett.30.884}{{\em Phys.
  Rev. Lett.} {\bfseries 30} (1973) 884--886}.

\bibitem{deRham:2012fw}
C.~de~Rham, A.~J. Tolley, and D.~H. Wesley, ``{Vainshtein Mechanism in Binary
  Pulsars},'' \href{http://dx.doi.org/10.1103/PhysRevD.87.044025}{{\em Phys.
  Rev. D} {\bfseries 87} no.~4, (2013) 044025},
  \href{http://arxiv.org/abs/1208.0580}{{\ttfamily arXiv:1208.0580 [gr-qc]}}.

\bibitem{Chu:2012kz}
Y.-Z. Chu and M.~Trodden, ``{Retarded Green\textquoteright{}s function of a
  Vainshtein system and Galileon waves},''
  \href{http://dx.doi.org/10.1103/PhysRevD.87.024011}{{\em Phys. Rev. D}
  {\bfseries 87} no.~2, (2013) 024011},
  \href{http://arxiv.org/abs/1210.6651}{{\ttfamily arXiv:1210.6651
  [astro-ph.CO]}}.

\bibitem{deRham:2012fg}
C.~de~Rham, A.~Matas, and A.~J. Tolley, ``{Galileon Radiation from Binary
  Systems},'' \href{http://dx.doi.org/10.1103/PhysRevD.87.064024}{{\em Phys.
  Rev. D} {\bfseries 87} no.~6, (2013) 064024},
  \href{http://arxiv.org/abs/1212.5212}{{\ttfamily arXiv:1212.5212 [hep-th]}}.

\bibitem{Joyce:2014kja}
A.~Joyce, B.~Jain, J.~Khoury, and M.~Trodden, ``{Beyond the Cosmological
  Standard Model},''
  \href{http://dx.doi.org/10.1016/j.physrep.2014.12.002}{{\em Phys. Rept.}
  {\bfseries 568} (2015) 1--98},
  \href{http://arxiv.org/abs/1407.0059}{{\ttfamily arXiv:1407.0059
  [astro-ph.CO]}}.

\bibitem{Dar:2018dra}
F.~Dar, C.~De~Rham, J.~T. Deskins, J.~T. Giblin, and A.~J. Tolley, ``{Scalar
  Gravitational Radiation from Binaries: Vainshtein Mechanism in Time-dependent
  Systems},'' \href{http://dx.doi.org/10.1088/1361-6382/aaf5e8}{{\em Class.
  Quant. Grav.} {\bfseries 36} no.~2, (2019) 025008},
  \href{http://arxiv.org/abs/1808.02165}{{\ttfamily arXiv:1808.02165
  [hep-th]}}.

\bibitem{Hellings:1983fr}
R.~W. Hellings and G.~S. Downs, ``{Upper limits on the isotropic gravitational
  radiation background from pulsar timing analysis},''
  \href{http://dx.doi.org/10.1086/183954}{{\em Astrophys. J. Lett.} {\bfseries
  265} (1983) L39--L42}.

\bibitem{Gair:2014rwa}
J.~Gair, J.~D. Romano, S.~Taylor, and C.~M.~F. Mingarelli, ``{Mapping
  gravitational-wave backgrounds using methods from CMB analysis: Application
  to pulsar timing arrays},''
  \href{http://dx.doi.org/10.1103/PhysRevD.90.082001}{{\em Phys. Rev. D}
  {\bfseries 90} (2014) 082001},
  \href{http://arxiv.org/abs/1406.4664}{{\ttfamily arXiv:1406.4664 [gr-qc]}}.

\bibitem{Allen:2022dzg}
B.~Allen, ``{Variance of the Hellings-Downs correlation},''
  \href{http://dx.doi.org/10.1103/PhysRevD.107.043018}{{\em Phys. Rev. D}
  {\bfseries 107} no.~4, (2023) 043018},
  \href{http://arxiv.org/abs/2205.05637}{{\ttfamily arXiv:2205.05637 [gr-qc]}}.

\bibitem{Allen:2022ksj}
B.~Allen and J.~D. Romano, ``{Hellings and Downs correlation of an arbitrary
  set of pulsars},'' \href{http://dx.doi.org/10.1103/PhysRevD.108.043026}{{\em
  Phys. Rev. D} {\bfseries 108} no.~4, (2023) 043026},
  \href{http://arxiv.org/abs/2208.07230}{{\ttfamily arXiv:2208.07230 [gr-qc]}}.

\bibitem{Mingarelli:2014xfa}
C.~M.~F. Mingarelli and T.~Sidery, ``{Effect of small interpulsar distances in
  stochastic gravitational wave background searches with pulsar timing
  arrays},'' \href{http://dx.doi.org/10.1103/PhysRevD.90.062011}{{\em Phys.
  Rev. D} {\bfseries 90} no.~6, (2014) 062011},
  \href{http://arxiv.org/abs/1408.6840}{{\ttfamily arXiv:1408.6840
  [astro-ph.HE]}}.

\bibitem{Domenech:2024pow}
G.~Dom\`enech and A.~Tsabodimos, ``{Finite distance effects on the
  Hellings-Downs curve in modified gravity},''
  \href{http://arxiv.org/abs/2407.21567}{{\ttfamily arXiv:2407.21567 [gr-qc]}}.

\end{thebibliography}\endgroup


\end{document}